\documentclass[12pt,preprint]{aastex}

\shorttitle{LCBGs in Galaxy Clusters}
\shortauthors{Crawford et al.}

\begin{document}

\title{Luminous Compact Blue Galaxies in Intermediate Redshift Galaxy
Clusters: A Significant But Extreme Butcher-Oemler Population}

\author{S. M. Crawford\footnote{Email: crawford@astro.wisc.edu}, 
M. A. Bershady, A. D. Glenn, and J. G. Hoessel}\affil{Washburn
Observatory, U. Wisconsin - Madison, 475 N. Charter St., Madison, WI
53706}

\begin{abstract}

We identify a population of Luminous Compact Blue Galaxies (LCBGs) in
two galaxy clusters: MS0451.6-0305 ($z=0.54$) and Cl1604+4304
($z=0.9$). LCBGs are identified via photometric characteristics and
photometric redshifts derived from broad and narrow band images taken
with the WIYN telescope and the Hubble Space Telescope. We analyze
their surface densities and clustering properties to find they compose
a statistically significant portion ($42\%$ and $53\%$) of the
Butcher-Oemler galaxies in both clusters, and their spatial
distributions are best characterized by a shell model. The enhancement
of the projected space-density of LCBGs with M$_B<-18.5$ in the
clusters relative to the field is 3-10 times higher than the BO
population as a whole, but 2 times lower than the red population,
except in the core where LCBGs are absent.  Assuming some fading, a
natural descendant would be small, low-luminosity galaxies found
preferentially in today's clusters, such as dEs.

\end{abstract}


\section{Introduction}

The number of blue, star forming galaxies increases in all
environments at intermediate redshifts ($0.3 < z < 1.0$).  In the
field, there is a dramatic rise in the space-density of
luminous\footnote{We adopt H$_{0}=70$ km s$^{-1}$ Mpc$^{-1}$,
$\Omega_{Matter} = 0.3$, and $\Omega_{\Lambda} = 0.7$.}  ($M_B\sim
-20$), compact ($R_e \sim 2 $kpc), and blue ($B-V \sim 0.35$) galaxies
known as LCBGs (Koo et al. 1994; Guzm\'{a}n et al. 1997). These
galaxies produce stars at such a tremendous rate ($1-40$ M$_{\odot}$
yr$^{-1}$, Hammer et al. 2001) they provide a substantial fraction of
the star formation in the universe at $0.4<z<1$ (Guzm\'{a}n et
al. 1997).  In clusters, Butcher \& Oemler (1978, 1984; hereafter BO)
claimed the fraction of blue galaxies increases with redshift. Blue
cluster galaxies have been classified as a mix of normal galaxies
absent in local clusters, morphologically disturbed, and star forming
galaxies (Oemler, Dressler, \& Butcher 1997). Recent studies indicate
star forming galaxies in intermediate redshift clusters are typically
small, disk-like (de Propris et al. 2002; Lotz, Martin, \& Ferguson
2003; Finn, Zaritsky, \& McCarthy 2003) and falling into the cluster
(Balogh, Navarro, \& Morris 2001; Homeier et al. 2004; Tran et
al. 2005) in groups and clumps (Kodama et al. 2002).  While several
studies exist for LCBGs in the field and Koo et al. (1997) have
identified a handful of LCBGs in CL0024, no thorough census of LCBGs
in clusters has been completed.

Field LCBGs and cluster star-forming galaxies have been proposed
as the progenitors of dwarf elliptical (dE) galaxies (Koo et al. 1994;
Guzm\'{a}n et al. 1997; Koo et al. 1997; Martin et al. 2000) or
low-mass S0's (Tran et al. 2005). The line widths and physical sizes
of field and cluster LCBGs are consistent with those of dwarf
elliptical galaxies (or field ``dEs'' like NGC205). Recent bursts
inferred from the stellar histories of local dwarf ellipticals
(Grebel, Gallagher, \& Harbeck 2004) and in nearby clusters (Poggianti
et al. 2001; Conselice, Gallagher, \& Wyse 2001) support a fading
scenario. LCBGs are viable candidates to explain, and eventually fill
in, the missing faint red sequence of galaxies seen at $z\sim0.75$ (de
Lucia et al. 2004, Goto et al. 2005). However, the high masses (P97),
high metallicities (Kobulnicky \& Zaritsky 1999), large extinctions
(Hammer et al. 2001), and centrally concentrated star-bursts (Barton
and van Zee 2001) seen in some LCBGs make them plausible candidates to
be a burst phase of more massive spiral galaxies. Both dE's and more
massive bulges are accreted populations in local clusters (Conselice
et al. 2001; Biviano et al. 2002), but they have very different
morphology-density relationships (Ferguson \& Bignelli
1994). Understanding the prevalance and distribution of LCBGs in
clusters should constrain their role as a progenitor population.

In this Letter we measure the density and clustering properties of
LCBGs in two rich clusters (Table 1): \objectname{MS
0451.6-0305} and \objectname{Cl1604+4304}. MS0451 is an incredibly
rich, X-ray bright cluster (Ellingson et al. 1998; Donahue et
al. 2003). Cl1604+4304 is part of a super-cluster complex (Postman,
Lubin, \& Oke 2001, Lubin, Mulchaey, \& Postman 2004, Gal \& Lubin
2005), but is not as x-ray luminous. The cluster redshifts span an
epoch where the dynamical mass of field LCBGs changes rapidly (P97),
and are sufficiently disparate to permit the derivation of
complementary field samples using the same data.

\section{Observations and Analysis}

Observations were obtained with the WIYN \footnote{The
WIYN Observatory is a joint facility of the University of
Wisconsin-Madison, Indiana University, Yale University, and the
National Optical Astronomy Observatories.} 3.5m telescope's
Mini-Mosaic Camera ($0.14''$ per pixel and $9.6' \times 9.6'$ field of
view) and augmented with archival HST WFPC2 and ACS images for both
clusters, reduced via the standard HST reduction pipeline.  MS0451 is
sampled by images in the F775W, F814W, and F850LP bandpasses; Cl1604
is imaged in F606W and F814W. Harris UBRI, Gunn z, and two narrow-band
filters \footnote{On-band filters were custom-made narrow-band
($\lambda / \Delta \lambda \sim 70$) filters sampling rest-frame
[OII]$\lambda3727$\AA .  Off-band filters were NOAO filter KP1582 ($
\lambda / \Delta \lambda \sim 20$) for MS0451 and another custom
filter ($ \lambda / \Delta \lambda \sim 70$) for Cl1604.} were
obtained at WIYN between 1999 October and 2004 June. We use data from
nights with good transparency and seeing (FWHM $\sim
0.85^{+0.45}_{-0.35}$ arcsec).

Reduced Mini-Mo images are flat to within $1\%$ of their initial sky
values.  We created deep mosaics by combining only high quality data
weighted by the ratio of the flux from an average star to the square
root of the sky deviation and seeing for that image (Bershady,
Lowenthal, \& Koo 1998). Data were calibrated through: (1)
spectro-photometric standard stars (Massey et al. 1988) and Landolt
(1992) photometric stars observed during the WIYN 3.5m runs; (2)
observations of Landolt (1992) standards and cluster fields at the
WIYN 0.9m telescope; (3) comparisons to the HST WFPC2 and ACS
observations; (4) comparison of the observed stellar locus to that
derived for our filter set from the Gunn-Stryker catalog (Gunn \&
Stryker 1983). Through these methods, we estimate relative and
absolute calibration uncertainties are below $2\%$, assuming no large
metallicity differences exist between the Gunn-Stryker catalog and our
field stars.

Object detection was performed on the sum of the UBRIz images using
SExtractor (Bertin \& Arnouts 1996) with the criterion that objects
contained $>20$ contiguous pixels above $3 \sigma$ of the sky noise.
Detection completeness was determined via Monte Carlo simulations
reinserting real objects back into the images. Total magnitudes and
half-light radii are determined from light-profile curves of growth.
Total magnitudes are set to the flux within an aperture which is a
multiple of the $\eta=0.2$ radius ($\eta$ is the Petrosian ratio as
defined by Kron 1995). The multiplier is determined from the
light-concentration parameter, $C_{2080}$ (Bershady, Jangren, \&
Conselice al. 2000).  With this ``tailored'' aperture, total
magnitudes are measured to an accuracy $1\%$ regardless of profile
shape (cf., only 80\% of the light is enclosed where $\eta=0.2$ for an
$r^{1/4}$ profil; Graham et al. 2005).  Colors were determined within
seeing-matched aperture of radius $1.5\times$FWHM in each
image. Photometric precision is 0.1 mag at $R=24.25,25$ for total and
seeing-matched apertures, respectively. Random errors for total
magnitudes depend on the light-profile shape; $r^{1/4}$-law profile
errors are larger than the values quoted here for exponential
profiles.

Photometric redshifts are determined through a method similar to
Csabai et al. (2003). We convolved a standard set of model and
observed galaxy templates with our filters to produce a
template-redshift grid for $0<z<5$ and $\Delta z=0.01$. Both fields
have close to 100 galaxies with spectroscopic redshifts (Ellingson et
al. 1998; Postman et al. 1998). For each field we match galaxies with
high-quality spectroscopic redshifts and good photometry with points
on the template-redshift grid, and correct the grid for differences
between the simulated and measured colors. The trained grids yield
photometric redshifts for every object with a precision $\sigma_z <
0.05$ (blue objects) and 0.03 (red objects) to $z=1.0$ and $S/N > 10$.

Absolute magnitudes, radii, rest-frame surface-brightness and colors
were calculated for all objects using photometric or spectroscopic
redshifts. K-correction calculations adopt method 4 of Bershady
(1995). Half-light radii were measured in the band closest to
rest-frame B, and corrected for PSF effects by quadrature subtraction
of the stellar half-light radius.  Surface brightnesses in the WIYN
images were corrected for seeing using measurements from the
overlapping regions with the HST data.  WIYN-based surface
brightnesses have a 1.5 mag dispersion due to seeing-correction
effects on the measured radii.

\section{Identification of LCBGs}

We define LCBGS here as ``enthusiastic'' star forming galaxies.  For
these purposes, we define LCBGs as having the following properties:
$(B-V)_o < 0.5$ and $\bar{\mu}_e(B) < 21$ mag arcsec$^{-2}$, where
$\bar{\mu}_e$ is the rest-frame B-band average surface brightness
within the half-light radius. A galaxy with $L_{bol} = 10^9 L_{\odot}
$ and a constant star formation rate will have $(B-V)_o < 0.4$ and
$\bar{\mu_e}(B) < 20.5$ for ages more than a few $\times$ 10$^8$ yr. A
moderate amount of extinction, e.g. $E(B-V)=0.1$, will leave the
galaxy with the above parameters.  Anything bluer or brighter will be
more than an enthusiastic star former.  The selection region for LCBGs
is plotted in Figure 1.  This region is mostly devoid of objects in
local surveys (e.g., Werk et al. 2004, Garland et al. 2004) and is
purposely constructed to identify actively star forming galaxies that
are extreme compared to the local universe in clusters or the field.
For comparison with other LCBG samples and to differentiate them from
``dwarf'' galaxies (at least in luminosity), we also require LCBGs to
have $M_B < -18.5$.

Galaxies were selected from the ground-based data set.  We estimate,
based on extant HST data, this selection misses $\sim5\%$ of the
bona-fide LCBGs ( primarily due to the error in the size measurement
), while introducing the same percentage of false classifications.  We
identified ``cluster'' galaxies in our sample-- including LCBGs -- as
having a photometric redshift within $\pm0.1$ of the cluster's
redshift and a reasonable probability to be at the cluster redshift
based on the individual photometric error and measurements from the
training set (Brunner and Lubin 2000).  Within 1 Mpc of the clusters'
centers, we find 41 candidate LCBGs.  Four objects are confirmed as
cluster members through spectroscopic redshifts, and 23 are confirmed
by having strong emission (equivalent width greater than 10 \AA ) in
the ``on'' narrow-band image, which samples rest-frame [OII]$\lambda
3727$.  We have 3 objects that are identified as cluster LCBGs, but
have spectroscopic redshifts that place them outside of the cluster
(still within a redshift of 0.1 of the cluster) and weak [OII]
emission.  The remaining 10 objects require spectroscopic follow-up to
confirm cluster membership.
 
We select two other groups of luminous (M$_B<-18.5$) cluster galaxies
in our data, red and blue, to compare to the LCBG population.  We fit
the color-magnitude relationship in both clusters, splitting the
population according to the classic definition of the Butcher-Oemler
effect (Butcher \& Oemler 1978, 1984): $|\Delta(B-V)_0| > 0.2$ are
``BO'' galaxies, which include the LCBG population.  The remainder are
the red cluster sequence.  We measure $f_b=0.22\pm0.05$ within $R=0.5$
Mpc, rising to $0.33 \pm 0.04$ at $R=1.5$ Mpc in MS0451, in good
agreement with Ellingson et al. (2001) and de Propris et al. (2003).
For Cl1604, we measure $f_b = 0.5\pm0.13$ within $R=0.5$ Mpc, and
rising to $0.63 \pm 0.07$ by $R=1.5$ Mpc, as compared to the values of
$f_b=0.8$ as measured by Rakos \& Schombert (1995).

\section{The Number Density and Distribution of Cluster LCBGs}

We plot the surface density of galaxies in Figure 2 as a function of
radius from the cluster center (defined as the brightest cluster
galaxy). The data extend to a radius where our completeness is still
uniform in the WIYN images. Two basic results emerge. (1) All
populations clearly show evidence of clustering. (2) LCBGs form a
statistically significant population in both clusters: $14\pm4\%$ and
$34\pm9\%$ of the total population at $R_{200}$ and M$_B<-18.5$ (for
MS0451 and Cl1604, respectively), and $42\pm11\%$ and $53\pm14\%$ of
the BO galaxies in each cluster. (Errors are likely non-Gaussian, and
arise from cluster-membership uncertainties.)  For comparison, we
estimate LCBGs constitute approximately $8\%$ and $26\%$ of all field
galaxies blue enough to be classified as ``BO'' at $z=0.5$ and 0.9.

We model the projected distribution of the LCBG, BO and red cluster
populations with a King profile (King 1972) and two spherical-shell
density profiles. The King radius is set to the scale radius of the
X-ray profile for each cluster (Donahue et al. 2003, Lubin et
al. 2004). Spherical-shell models are empty inside of the X-ray scale
radius (``shell'') or half this value (``half-shell''), and then both
decline as a King profile with a 1 Mpc core radius. The King model
fits all of the populations well in a $\chi^2$ sense.  The red
population is only fit well by the King model; the BO populations are
best fit by the half-shell model; and LCBGs in MS0451 are best fit by
the shell model, whereas the Cl1604 LCBGs are best fit by the
half-shell model.  If LCBGs are distributed as a King profile, there
is a $95\%$ probability we would detect $\geq 1$ LCBG in the inner
0.15 Mpc region of MS0451. Therefore, LCBGs do not appear to exist in
central cluster regions -- in agreement with Homeier et al. (2005)
findings for star forming galaxies in Cl0152 (z=0.84).

We compare the surface-density of cluster objects to the same objects
in the field.  Red and BO galaxy field densities are calculated from
the luminosity function of the DEEP2 data (Faber et al. 2005). With
our selection criteria, the field LCBG surface density derived from
the P97 sample yields 1.2 and 5.44 $Mpc^{-2}$ at $z=0.53$ and
$z=0.90$, respectively, after applying corrections due to their
selection effects. We also use foreground LCBGs in the Cl1604 field to
estimate the field density of LCBGs at $z \sim 0.5$, and vice versa
for MS0451, to find $1.9\pm0.8$ and $4.7\pm1.2$ $Mpc^{-2}$ at $z=0.53$
and $z=0.90$ for objects with $M_B < -18.5$. Field densities are
calculated in the same redshift bin-size and manner as the cluster
samples. We couch our comparison between the two objects in terms of
``enhancement:'' the surface-density ratio of cluster to field.  The
enhancement as a function of cluster-galaxy surface density (Figure 3)
shows red galaxies behave qualitatively as expected according to the
morphology-density relationship (Dressler et al. 1997). The BO
galaxies show only a modest enhancement and trend with surface
density. LCBGs display a larger enhancement, increasing with luminous
(M$_B<-18.5$) galaxy surface density, but dropping precipitously at
surface densities above 150 Mpc$^{-2}$.

\section{Discussion}

We find LCBGs in two intermediate redshift clusters compose a
significant fraction of the BO populations despite the clusters'
different redshifts, environments (x-ray luminosities, and projected
densities), and relative blue fractions.  At lower redshift and
greater ``richness'' (MS0451), there is clear segregation in the
surface-densities of sub-populations with red:BO:LCBG = 26.6:12.9:5.4
Mpc$^{-2}$ within R$_{200}$. In Cl1604 the different populations have
comparable densities (red:BO:LCBG =8.9:15.8:8.5 Mpc$^{-2}$). At
luminous-galaxy surface-densities of 100 Mpc$^{-2}$, the enhancement
of these different sub-populations relative to the field are a factor
of 2-4 for BO galaxies, 8-20 for LCBGs, and 30-50 for red galaxies in
both clusters. This is suggestive of LCBGs as progenitors of
populations found preferentially in clusters today. Because LCBGs show
significant enhancement variations between clusters and with
surface-density, a secure interpretation of LCBGs as a progenitor
population awaits better sampling of environment and redshift. An
analysis of 10 clusters between $0.3<z<1.0$ is forthcoming. Here we
suggest that since they appear to be physically small, if LCBGs fade,
dEs and low-mass S0s would be one plausible remnant population.

Despite their similar spatial distribution, LCBGs are an extreme
sub-component of the BO population in terms of their cluster
enhancement, color, and surface-brightness. If LCBGs are on
predominantly radial orbits, their cluster-shell enhancement indicates
initial in-fall -- a spectroscopically testable claim. In this
scenario, their star-bursts are plausibly triggered through galaxy
interactions in the cluster periphery, where densities are enhanced
but interaction times still long, or via interactions with the
intercluster medium (ICM). The former supposition is testable via
angular correlation measurements of large samples. A correlation
between the X-ray emission and position of the LCBGs would support an
ICM-driven trigger for the star-bursts.

\acknowledgments

We thank the anonymous referee for helpful comments. Research was
supported by STScI/AR-9917, NSF/AST-0307417, and a Wisconsin Space
Grant. Some of the data presented in this paper were obtained from
STScI/MAST.

\clearpage

\begin{deluxetable}{lrrrrrrr}
\tabletypesize{\scriptsize}
\tablewidth{0pt}
\tablecaption{Cluster Properties}
\tablehead{
\colhead{Cluster} & 
\colhead{RA} & 
\colhead{DEC} & 
\colhead{z} & 
\colhead{$\sigma$} & 
\colhead{$L_x$\tablenotemark{a}} & 
\colhead{$R_c$} & 
\colhead{$R_{200}$\tablenotemark{b}} 
\\
\colhead{} & 
\colhead{J2000} &  
\colhead{J2000} & 
\colhead{}& 
\colhead{(km s$^{-1}$)} & 
\colhead{(erg s$^{-1}$)} &
\colhead{(Mpc)} & \colhead{(Mpc)} 
}
\startdata
MS0451 & 04:54:10.8 & -03:00:56 & 0.538 & 1354 &1.4$\times 10^{45}$ & 0.18 & 1.64 \\
Cl1604 & 16:04:18.2 & +43:04:38 & 0.90 & 982 &8.6$\times 10^{43}$& 0.14 &  1.48 \\
\enddata
\tablenotetext{a}{$L_x$ is measured in the 0.3-3.5 keV and 0.5-2.0 keV
bands, respectively, for MS0451 and Cl1604.}

\tablenotetext{b}{$R_{200}$ is the radius where the cluster density
is 200 times the critical density (Finn et al. 2004).}

\end{deluxetable}

\begin{figure}[hbt]
\epsscale{0.90}
\plotone{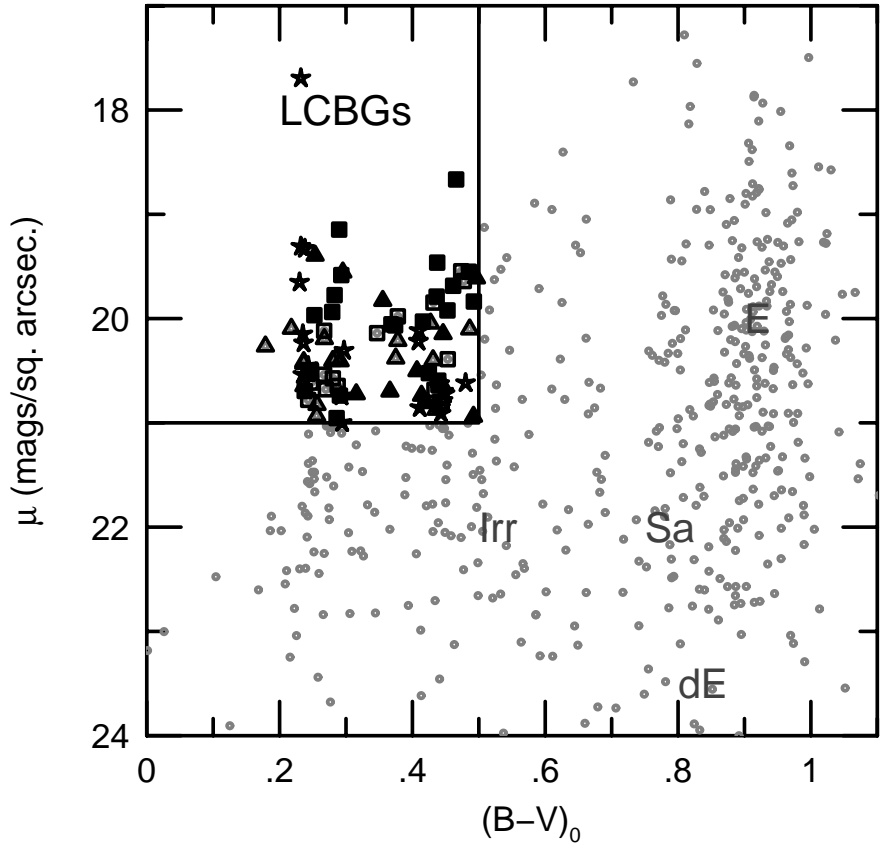}
\caption{Rest-frame surface-brightness vs. color  for cluster LCBGs (filled
symbols) and field LCBGs (open symbols) found in the HST images of
MS0451 and Cl1604, and field LCBGs from P97 (stars).  Other
intermediate redshift galaxies in our fields are grey points.  Lines
mark the selection criteria for LCBGs used here.  Labels indicate z=0
galaxy types.}
\end{figure}

\begin{figure}[hbt]
\epsscale{0.70}
\plotone{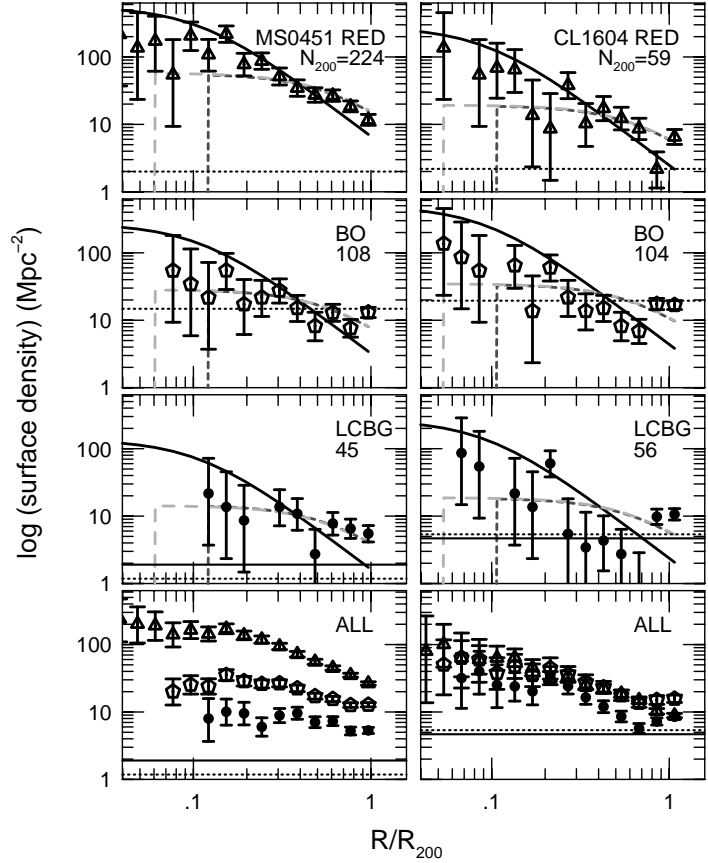}
\caption{Surface-densities for cluster
galaxies. The top six panels are the differential distributions of red
galaxies (triangles), BO galaxies (pentagons), and LCBGs (circles) for
MS0451 and Cl1604.  Model curves (see text) are for a King profile
(solid curve) and shell models (dashed) based on the X-ray light
profile. Dotted lines represent field densities for each class at the
cluster redshift (see text). Solid lines are the LCBG field density as
measured in our ``off'' image. The number, $N_{200}$, is the number of
objects at $R_{200}$ in each respective class. Bottom panels show mean
surface-densities.}
\end{figure}

\begin{figure}[hbt]
\epsscale{1.0}
\plotone{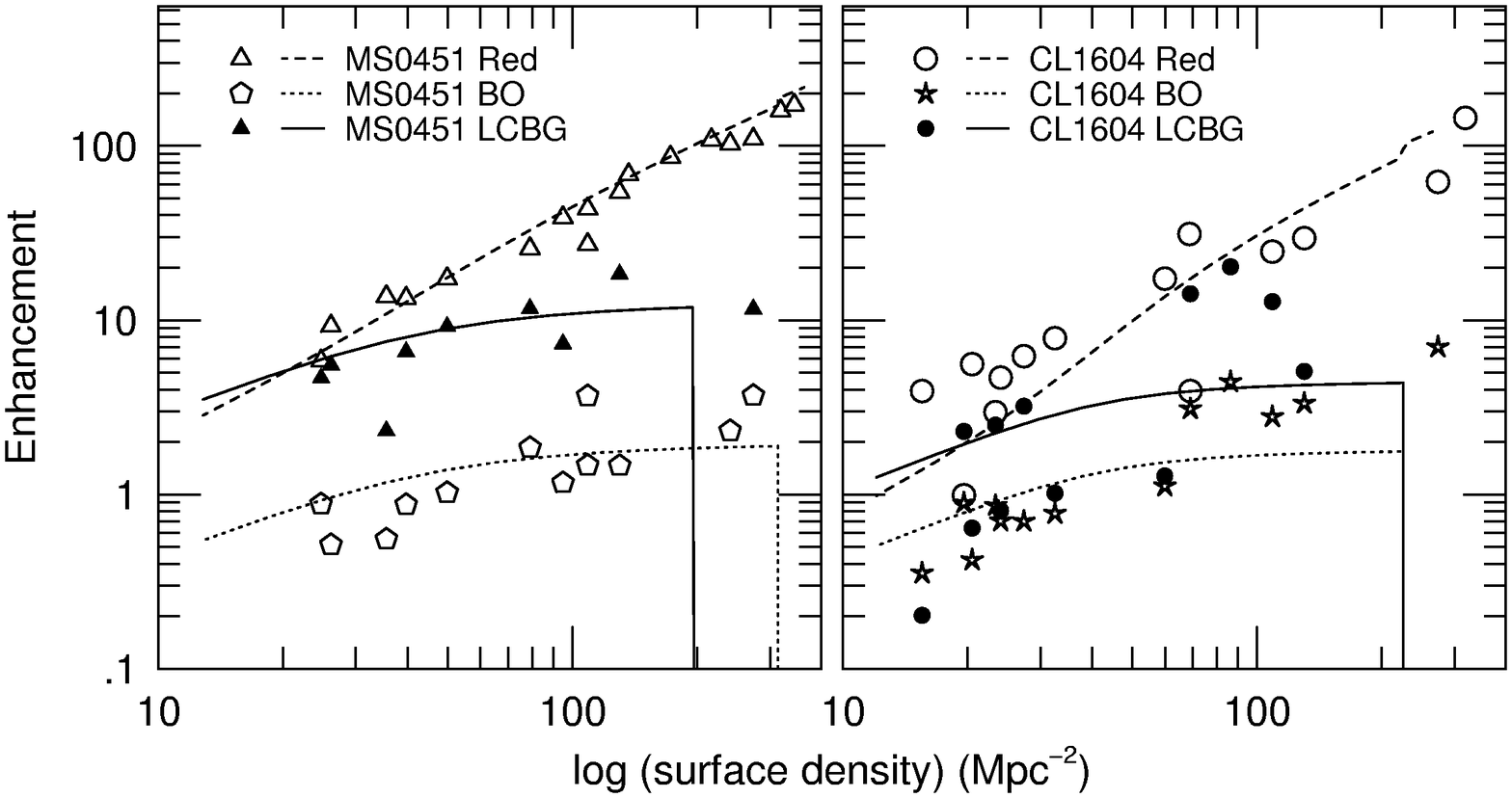}
\caption{Cluster galaxy enhancement relative to the field as a
function of cluster surface density (for galaxies with M$_B<-18.5$
within circular annuli about the cluster center). The lines are the
best fit models to each population.}
\end{figure}


\begin{references}

\reference {} {Balogh, M.~L., Navarro, J.~F., \& Morris, S.~L.\ 2000, \apj, 540, 113} 
\reference {} {Barton, E.~J., \& van Zee, L.\ 2001, \apjl, 550, L35 }
\reference{} {Bershady, M.~A.\ 1995, \aj, 109, 87 }
\reference{} { Bershady, M. A., Lowenthal, J. D. and Koo, D. C. 1998, \apj, 505, 50 }
\reference {} {Bershady, M.~A., Jangren, A., \& Conselice, C.~J.\ 2000, \aj, 119, 2645 }
\reference {} {Bertin, E.~\& Arnouts, S.\ 1996, \aaps, 117, 393}
\reference {} {Biviano, A., Katgert, P., Thomas, T., \& Adami, C.\ 2002, \aap, 387, 8} 
\reference {} {Brunner, R.~J., \& Lubin, L.~M.\ 2000, \aj, 120, 2851} 
\reference{} {Butcher, H.~\& Oemler, A.\ 1978, \apj, 226, 559 }
\reference{} {Butcher, H.~\& Oemler, A.\ 1984, \apj, 285, 426 }
\reference {} {Conselice, C.~J., Gallagher, J.~S., \& Wyse, R.~F.~G.\ 2001, \apj, 559, 791 }
\reference {} {Csabai, I., et al.\ 2003, \aj, 125, 580 }
\reference {} {de Lucia, G, et al. \ 2004, \apj, 610, L77 } 
\reference {} {de Propris, R., Stanford, S.~A., Eisenhardt, P.~R., \& Dickinson, M.\ 2003, \apj, 598, 20}
\reference {} {Dressler, A.~et al.\ 1997, \apj, 490, 577 }
\reference {} {Donahue, M., Gaskin,J.~A., Patel, S.~K., Joy, M., Clowe, D., \& Hughes, J.~P.\ 2003, \apj, 598, 190 }
\reference {} {Ellingson, E., Yee, H.~K.~C., Abraham, R.~G., Morris, S.~L., \& Carlberg, R.~G.\ 1998, \apjs, 116, 247 }
\reference {} {Ellingson, E., Lin, H., Yee, H.~K.~C., \& Carlberg, R.~G.\ 2001, \apj, 547, 609 }
\reference {} {Faber, S. M., et al. \ 2005. astro-ph/0506044 }
\reference {} {Ferguson, H.~C.~\& Binggeli, B.\ 1994, \aapr, 6, 67 }
\reference {} {Finn, R.~A., Zaritsky, D., \& McCarthy, D.~W.\ 2004, \apj, 604, 141 }
\reference {} {Gal, R.~R., \& Lubin,L.~M.\ 2004, \apjl, 607, L1} 
\reference {} {Garland, C.~A., Pisano, D.~J., Williams, J.~P., Guzm{\' a}n, R., \& Castander, F.~J.\ 2004, \apj,615, 689 }
\reference {} {Goto, T, et al. \ 2005, \apj, 621, 188}
\reference {} {Grebel, E.~K., Gallagher, J.~S., \& Harbeck, D.\ 2003, \aj, 125, 1926}
\reference {} {Graham, A.~W., et al. \ 2005, astro-ph/0504287}
\reference {} {Gunn, J.~E.~\& Stryker, L.~L.\ 1983, \apjs, 52, 121 }   
\reference {} {Guzman, R., Gallego, J., Koo, D.~C., Phillips, A.~C., Lowenthal, J.~D., Faber, S.~M., Illingworth, G.~D., \& Vogt, N.~P.\ 1997, \apj, 489, 559 }
\reference {} {Hammer, F.~, Gruel, N., Thuan, T.~X., Flores, H., \& Infante, L.\ 2001, \apj, 550, 570}
\reference {} {Homeier, N.~L., et al.\ 2005, \apj, 621, 651 } 
\reference {} {King, I.~R.\ 1972, \apjl, 174, L123 }
\reference {} {Kobulnicky, H.~A.~\& Zaritsky, D.\ 1999, \apj, 511, 118.}
\reference {} {Kodama, T., Smail, I., Nakata, F., Okamura, S., \& Bower, R.~G.\ 2001, \apjl, 562, L9 } 
\reference {} {Koo, D.~C., Bershady, M.~A., Wirth, G.~D., Stanford, S.~A., \& Majewski, S.~R.\ 1994, \apjl, 427, L9.}
\reference {} {Koo, D.~C., Guzman, R., Gallego, J., \& Wirth, G.~D.\ 1997, \apjl, 478, L49}
\reference {} {Kron, R. G. 1995, in The Deep Universe, ed. B. Bingeli \& R. Buser (New York: Springer), 233}
\reference {} {Landolt, A.~U.\ 1992, \aj, 104, 340 } 
\reference {} {Lotz, J.~M., Martin, C.~L., \& Ferguson, H.~C.\ 2003, \apj, 596, 143 }
\reference {} {Lubin, L.~M., Mulchaey, J.~S., \& Postman, M.\ 2004, \apjl, 601, L9 }
\reference {} {Massey, P., Strobel,\ 1988, \apj, 328, 315 } 
\reference {} {Martin, C.~L., Lotz, J., \& Ferguson, H.~C.\ 2000, \apj, 543, 97 }
\reference {} {Oemler, A.~J., Dressler, A., \& Butcher, H.~R.\ 1997, \apj, 474, 561 }
\reference {} {Phillips, A.~C., Guzman, R., Gallego, J., Koo, D.~C., Lowenthal, J.~D., Vogt, N.~P., Faber, S.~M., \& Illingworth, G.~D.\ 1997, \apj, 489, 543 (P97) }
\reference{} {Poggianti, B.~M., et al.\ 2001, \apj, 562, 689} 
\reference {} {Postman, M., Lubin, L.~M., \& Oke, J.~B.\ 2001, \aj, 122, 1125 }
\reference {} {Rakos, K.~D., \& Schombert, J.~M.\ 1995, \apj, 439, 47}
\reference{} {Tran, K.~H., van Dokkum, P., Illingworth, G.~D., Kelson, D., Gonzalez, A., \& Franx, M.\ 2005, \apj, 619, 134 }
\reference {} {Werk, J.~K., Jangren, A., \& Salzer, J.~J.\ 2004, \apj, 617, 100 }

\end{references}
\end{document}